\documentclass[twocolumn,secnumarabic,amssymb, nobibnotes, aps, prd]{revtex4}
\usepackage{graphicx}

\begin{document}
\title{Discrete spectrum of measured parameters of a superconductor nanostructure.}
\author{A.V.  Nikulov}
\affiliation{Institute of Microelectronics Technology and High Purity Materials, Russian Academy of Sciences, 142432 Chernogolovka, Moscow District, RUSSIA.} 
\begin{abstract} The discreteness of permitted state spectrum postulated on atomic level can be macroscopic in nanostructures and larger structures because of macroscopic quantum phenomena such as superconductivity. The change by jump of measured parameters because of the macroscopic discreteness may be used for some applications. A device is proposed measured parameters of which can change by jump at a weak change of external parameters. The devise consists of two superconductor loop connected with two Josephson junctions. The macroscopic parameter - maximum value of the super-current through the two Josephson junctions - can change with the quantum number determining macroscopic angular momentum of superconducting pairs in one of the two loops.
 \end{abstract}

\maketitle

\narrowtext

\section*{Introduction}

The distinctive feature of quantum mechanics is discreteness. Its scale on the atomic level is determined with the Planck's constant $\hbar $. For example, the discrete values of measured projection of angular momentum of atom and spin differ on $\hbar $. Discreteness in nanostrucrures may have much larger scale because of macroscopic quantum phenomena such as superconductivity. According to the Bohr's quantization $pr = n\hbar $ the difference of angular momentum $M_{p} = pr$  between adjacent permitted states, $n + 1$ and $n$, of a free electron in a nano-ring with a radius $r > 1 nm$ equals $\hbar $ as well as in atom. The energy difference $E_{n+1} - E_{n} = mv_{n+1}^{2}/2 - mv_{n}^{2}/2 = (2n+1)\hbar ^{2}/2mr^{2}$ is much smaller than in atom. Because of the latter the persistent current, the quantum phenomenon observed because of the discreteness of spectrum, can be observed only at very low temperature in semiconductor and normal metal rings with realistic radius from $r = 10 \ nm$  [1] to $r = 800 \ nm$ [2]. This phenomena can not be practically observed in the ring with the radius $r > 10 \ \mu m$. In contrast to the semiconductor and normal metal rings, in superconductor ring with such radius the persistent current is observed in the whole temperature region corresponding to the superconducting state $T < T_{c}$ and even in some region above it $T > T_{c}$ [3]. The energy difference $E_{n+1} - E_{n}$ between adjacent permitted states in superconductor ring in $N_{s} $ times higher [4] then in the non-superconductor one because of the same $n\hbar $ angular momentum of all $N_{s} = Vn_{s} = \pi r^{2}sn_{s}$ superconducting pairs containing in the ring with the volume $V$,  the radius $r$ and the section area $s$.

The angular momentum difference between adjacent permitted states of superconductor rings with uniform section area $s$ increases also in $N_{s} $ times, $M_{p,n+1} - M_{p,n} = N_{s} \hbar $. The density of pairs $n_{s}$ in superconducting state $T < T_{c}$ is equal in order of value 1/2 electron density in the metal. Therefore the number $N_{s} $ is enormous in any ring with realistic size. The great discreteness may be useful for applications. Nanostructures may be more useful for some applications because of too great discreteness of macroscopic superconductor structures. In order to the discreteness can be used it should become apparent in measurable parameters. A device measurable parameters of which is strongly discrete is proposed in this work.  

\section {Discrete states of superconducting loop}
The enormous discreteness in superconductor is describe very well with the Ginzburg-Landau wave function $\Psi_{GL} =|\Psi _{GL}|\exp{i\phi }$. Where $|\Psi _{GL}|^{2} = n_{s}$ and $V|\Psi _{GL}|^{2} = Vn_{s} = N_{s} $ are the density and the total number of superconducting pairs in the ring. But $\hbar \bigtriangledown \phi = p = mv + qA$ is canonical momentum of single pair with the charge $q = 2e$. Because of the requirement that the complex pair wave function closed in the loop must be single-valued at any its point $\Psi_{GL} =|\Psi_{GL} |\exp{i\phi }= |\Psi _{GL}|\exp{i(\phi + 2\pi n )}$ the phase $\phi $ must change by integral multiples of $2\pi$  
$$\oint_{l} dl  \bigtriangledown \phi = n2\pi  \eqno{(1)}$$ 
following a complete turn along the path of integration $l$. Because of the quantization (1) and the relation $\oint_{l} dl  \hbar \bigtriangledown \phi = \oint_{l} dl p = m\oint_{l} dl v + q\Phi $ the minimum value of the permitted velocity $v = (2\pi\hbar/ml)(n - \Phi/\Phi_{0})$ of superconducting pairs in a loop is periodic function of magnetic field $B$ with the period  $B_{0} = \Phi_{0}/S$ corresponding to the flux quantum $\Phi _{0} = 2\pi \hbar /q \approx 2.07 \ 10^{-15} \ T m^{2}$  inside the loop with the area $S$. The two permitted states $n = k$ and $n = k+1$ have equal minimum energy $E_{k} = E_{k+1} = N_{s}mv^{2}/2 = N_{s} \pi^{2}\hbar^{2}/ml^{2}$ at $BS = \Phi = (k+0.5) \Phi _{0}$. Because of the great number of pairs $N_{s}$ the energy difference between these states becomes high $|E_{k+1} - E_{k}| > k_{B}T  $ at small deviation $\delta B = B - (k+0.5) \Phi _{0}/S$ of magnetic field value $B$ from $(k+0.5) \Phi _{0}/S$. The angular momentum and other parameters can change by jump at small variation $\delta B$ of the magnetic field $B$.   

\begin{figure}[]
\includegraphics{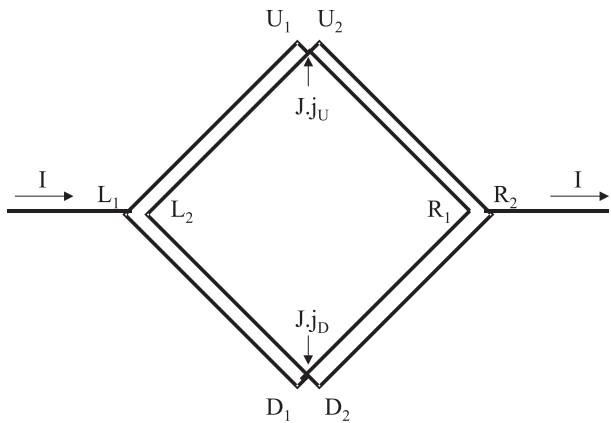}
\caption{\label{fig:epsart} A sketch of the device consisting of two superconductor loop connected with two Josephson junctions $J.j_{U}$ and $J.j_{D}$. The maximum value of the super-current $I_{s}$ changes by jump from $ I_{s}=2I_{c}$ to $I_{s} = 0$ at the change the the quantum number $n_{1}$ or $n_{2}$ in one of the loops. This jump can be observed as voltage $V = R(I - I_{s})$ jump at applying of an external current $I  < 2I_{c}$ as shown on the figure.}
\end{figure}

\section {Two superconductor loop connected with two Josephson junctions }
This change can be detected with help of the device, sketch of which is shown on Fig.1. The device consists of two superconductor loops $L_{1} - U_{1} - R_{1} - D_{1} - L_{1}$ and $L_{2} - U_{2} - R_{2} - D_{2} - L_{1}$ with equal area $S$. The velocity of superconducting pairs in the loops $v_{1} = (2\pi\hbar/ml)(n_{1} - \Phi/\Phi_{0})$ and $v_{2} = (2\pi\hbar/ml)(n_{2} - \Phi/\Phi_{0})$ depends on the quantum numbers $n_{1}$ and $n_{2}$, i.e. the phase $\phi $ change (1) at complete turn along each of the loops. The loops are connected with two Josephson junctions in two points, as it is shown on Fig.1. According to the current-phase relationship $I = I_{c}\sin (\Delta \phi )$ between the super-current $I_{s}$ through the Josephson junction and the phase difference $\Delta \phi$ between the junction boundaries the maximum value of the super-current 
$$I_{s} = I_{c}[\sin(\Delta \phi_{U}) + \sin(\Delta \phi_{D})]  \eqno{(2)}$$
through two Josephson junction should depend on the relation between the phase differences $\Delta \phi_{U}$ and $\Delta \phi_{D}$. According to the quantization relation (1) along the path of integration $L_{1} - U_{1} - R_{1} - D_{1} - L_{1}$, $L_{2} - U_{2} - R_{2} - D_{2} - L_{2}$ and  $L_{1} - U_{1} - J.j_{U} - U_{2} - R_{2} - D_{2} - J.j_{D} - D_{1} - L_{1}$ the value $\Delta \phi_{D} - \Delta \phi_{U} = \pi(n_{1} + n_{2})$ is determined only with the quantum numbers $n_{1}$ and $n_{2}$, at any external magnetic field $B$ and the equal area $S$ of the loops. The maximum value of the super-current (2) 
$$I_{s} = I_{c}[\sin(\Delta \phi_{U}) + \sin(\Delta \phi_{U}+\pi(n_{1} + n_{2}))]  \eqno{(3)}$$ 
should have discrete values equal $2I_{c}$ when the sum $n_{1} + n_{2}$ is an even number and $I_{s} = 0$ when $n_{1} + n_{2}$ is an odd number.

The persistent current $I_{p} = s2en_{s}v = (4sen_{s}\pi\hbar/ml)(n - \Phi/\Phi_{0})$ in the states $n = k$ and $n = k+1$ with minimum energy at $BS = \Phi = (k+0.5) \Phi _{0}$ has the same direction in the loops when the sum $n_{1} + n_{2}$ is an even number and the directions are opposite when $n_{1} + n_{2}$ is odd. The change of the persistent current direction, for example, in the first loop at the change from $n_{1} = k$ to $n_{1} = k+1$ can be detected as the voltage $V = R_{n}(I - I_{s})$ jump, at an external current $I  < 2I_{c}$ applied as shown on Fig.1. The voltage jump can mount up to the value $ R_{n}I_{c} = \pi \Delta /2e = \pi 1.76 k_{B}T_{c}/2e$ [5]. For example, the maximum jump can equal $ 2 \ mV$ in the case on niobium loops with the critical temperature $T_{c} = 9.2 \ K$.  

This jump may be observed at very weak variation $\delta B = B - (k+0.5) \Phi _{0}/S$  of the external magnetic field $B$. The energy $E_{n} = N_{s}mv_{n}^{2}/2 = I_{p,A}\Phi _{0} (n - \Phi /\Phi _{0})^{2}$ difference between adjacent permitted states, for example $n = 0$ and $n = 1$ at $BS = \Phi = 0.5 \Phi _{0}$, 
$$E_{1} - E_{0} = 2I_{p,A}S \delta B = 2I_{p,A}\Phi_{0}(S \delta B/\Phi_{0})  \eqno{(3)}$$  
increases with the loop area $S$ and the amplitude $I_{p,A} = 2sen_{s}\pi\hbar/ml$ of the oscillations $I_{p} = I_{p,A}2(n - \Phi /\Phi _{0})$. The value $\delta B > (E_{1} - E_{0})/2I_{p,A}S $ of the magnetic field variation, at which the probability $P_{n=1} =  e^{-(E_{1} - E_{0})/k_{B}T}/[1 + e^{-(E_{1} - E_{0})/k_{B}T} ]$ of the $n = 1$ state, for example, changes between 0 and 1, decreases with the increase of the loop sizes, because of the relation $I_{p,A} S \propto (s/l)l^{2} = sl = V$. This value is enough low already in a loop with nano-size. For example, the persistent current amplitude $I_{p,A} \approx 200 \ \mu A (1 - T/T_{c})$ observed in the aluminum ring with diameter $2r \approx  4 \ \mu m$ and circumference section $s \approx  1000 \ nm^{2}$ [6] corresponds to the value $2I_{p,A}\Phi_{0}/k_{B} \approx  30000 \ K $ at $T \approx  0.5 T_{c}$. The jump from $P_{n=1} = 0$ to $P_{n=1} = 1$ should by observed with variation $S \delta B > 0.0003 \Phi_{0} \approx 0.5 \ 10^{-18} \ T m^{2}$ and $\delta B > 5 \ 10^{-8} \ T $ at this value of the persistent current, $T \approx 10 \ K$ and $S = \pi r^{2} \approx 12 \ \mu m = 1.2 \ 10^{-11} \ m^{2}$.

The magnetic flux variation $S \delta B = \delta \Phi > 0.0007 \Phi_{0} = 10^{-18} \ T m^{2}$ can be induced with a current $I_{sw} > 2 \ 10^{-8} \ A = 0.02 \ \mu A$ circulating in an additional loop with  $2r = 4 \ \mu m$ having the inductance $L = 2 \ 10^{-11} \ H$ [5]. The value of the switching current $I_{sw}$ should not depend on the loop length $l$, because of the proportionality $L \propto  l$ and $ I_{p,A} \propto 1/l$. The loop with a small $l$ can be used as a bit, the two states $n = 0$ and $n = 1$ of which can be read in the way shown on Fig.1. The devise Fig.1 with a large loop length $l \gg  \pi 2r \approx  12 \ \mu m$ can be used for measurement of very weak magnetic field $B \approx  \delta B \propto 1/sl$. This very sensitive magnetometer has some advantages in comparison with the well known SQUID (Superconducting quantum interference devices) [7].   

The possibility of the voltage jump at the quantum number $n_{1}$, $n_{2}$ change in one of the loop of the devise shown on Fig.1 was corroborated experimentally in the work [8]. The voltage $V = R_{n}(I - I_{s})$ (or the resistance $V/I$ measured in [8]) changes by jump with magnetic field $B$ at the change of the persistent current direction, i.e. the $n_{1}$ number in one of the loop and returns to the initial value at the same change in the other loop. The quantum numbers $n_{1}$ and $n_{2}$ change in [8] at different magnetic field values because of different section $s$ of the loops. This circumstance may be used in the devise proposed in the present work.   

\section*{Acknowledgement}
This work has been supported by a grant "Possible applications of new mesoscopic quantum effects for making of element basis of quantum computer, nanoelectronics and micro-system technic" of the Fundamental Research Program of ITCS department of RAS and the  Russian Foundation of Basic Research grant 08-02-99042-r-ofi.

\end{document}